\documentclass[reprint,prb,twocolumn,showpacs,superscriptaddress,aps]{revtex4-1}
\usepackage[utf8]{inputenc}

\usepackage{graphicx}
\DeclareGraphicsExtensions{.png,.jpg,.eps}

\usepackage{float}
\usepackage{xcolor}
\usepackage{amsmath}
\usepackage{caption}
\usepackage{subcaption}
\usepackage[printwatermark]{xwatermark}
\usepackage{float}

\begin{document}

\title{Electronic structure and magnetic exchange interactions of Cr-based van der Waals ferromagnets. A comparative study between CrBr$_3$ and Cr$_2$Ge$_2$Te$_6$}

\author{Adolfo O. Fumega}
  \email{adolfo.otero.fumega@usc.es}
 \affiliation{Departamento de F\'{i}sica Aplicada,
  Universidade de Santiago de Compostela, E-15782 Campus Sur s/n,
  Santiago de Compostela, Spain}
\affiliation{Instituto de Investigaci\'{o}ns Tecnol\'{o}xicas,
  Universidade de Santiago de Compostela, E-15782 Campus Sur s/n,
  Santiago de Compostela, Spain}   
\author{S.~Blanco-Canosa}
\email{sblanco@dipc.org}
\affiliation{Donostia International Physics Center, DIPC, 20018 Donostia-San Sebastian, Basque Country, Spain}
\affiliation{IKERBASQUE, Basque Foundation for Science, 48013 Bilbao, Basque Country, Spain}
 \author{H. Babu-Vasili}
\affiliation{ALBA Synchrotron Light Source, Cerdanyola del Vall\`es, 08290 Barcelona, Catalonia, Spain}
 \author{Jian-Shi Zhou}
\affiliation{Texas Materials Institute, ETC 9.102, The University of Texas at Austin, Austin, Texas, 78712, USA.}
 \author{F. Rivadulla}
\affiliation{CIQUS, Centro de Investigaci\'{\o}n en Qu\'{\i}mica Biolóxica e Materiais Moleculares, Universidade de Santiago de Compostela, 15782-Santiago de Compostela, Spain.}
\author{Victor Pardo}
  \email{victor.pardo@usc.es}
\affiliation{Departamento de F\'{i}sica Aplicada,
  Universidade de Santiago de Compostela, E-15782 Campus Sur s/n,
  Santiago de Compostela, Spain}
\affiliation{Instituto de Investigaci\'{o}ns Tecnol\'{o}xicas,
  Universidade de Santiago de Compostela, E-15782 Campus Sur s/n,
  Santiago de Compostela, Spain}

\begin{abstract}

Low dimensional magnetism has been powerfully boosted as a promising candidate for numerous applications. The stability of the long-range magnetic order is directly dependent on the electronic structure and the relative strength of the competing magnetic exchange constants. Here, we report a comparative pressure-dependent theoretical and experimental study of the electronic structure and exchange interactions of two-dimensional ferromagnets CrBr$_3$ and Cr$_2$Ge$_2$Te$_6$. While CrBr$_3$ is found to be a Mott-Hubbard-like insulator, Cr$_2$Ge$_2$Te$_6$ shows a charge-transfer character due to the broader character of the Te 5p bands at the Fermi level. This different electronic behaviour is responsible of the robust insulating state of CrBr$_3$, in which the magnetic exchange constants evolve monotonically with pressure, and the proximity to a metal-insulator transition predicted for Cr$_2$Ge$_2$Te$_6$, which causes a non-monotonic evolution of its magnetic ordering temperature. We provide a microscopic understanding for the pressure evolution of the magnetic properties of the two systems.

\end{abstract}

\maketitle

\section{Introduction}

Two-dimensional (2D) materials have increased their academic and technological relevance continuously since the successful synthesis of graphene in 2004.\cite{graphene} Many other materials, like transition metal dichalcogenides,\cite{mos2} boron nitride (BN)\cite{bn} or phosphorene, have been profusely studied since then.\cite{phosphorene} Recently, long-range ferromagnetic (FM) order in several atomic-thick materials, such as FePS$_3$,\cite{feps3_2016} CrI$_3$ \cite{cri3_2015} and itinerant Fe$_3$GeTe$_2$ \cite{fegete} has been reported. The advent of long-range 2D ferromagnetism brings about new transport phenomena in two dimensions, like tunneling magnetoresistance\cite{Ghazaryan2018, tmr_lado} and  electrical switching of magnetic states,\cite{jarillo_nat_nano} promoting 2D ferromagnets as versatile platforms for engineering new quantum states and device functionalities. Besides, 2D materials can exhibit multitude of exotic properties when combined in heterostructures \cite{Cao2018}. 

The successful realization of a 2D magnet relies on the anisotropic exchange interaction (driven by spin-orbit coupling, strain, etc.) at the ultra-thin limit, that opens a gap in the spin excitation spectrum.  
In addition, in the low-dimensional limit with the absence of out-of-plane coupling, the interplay between the different exchange interaction paths is anticipated to become strongly renormalized. For instance, the isotropic in-plane magnetic coupling \textit{J}$_{in}$ in transition metal dichalcogenides (TMDs), di- and trihalides and phosphosulfides, is determined by the competition between metal-metal direct exchange and indirect exchange mediated by the anions.\cite{goody} Successive stacking of layers of 2D magnets, stabilized by van der Waals interactions, provides therefore a unique platform to study the evolution of the magnetic exchange interactions when going from the 2D to the 3D limit.

%WHAT DO YOU MEAN BY "STRONGLY RENORMALIZED" IN THE PREVIOUS PARAGRAPH?

\begin{figure*}[!ht]
  \centering
  \includegraphics[width=\textwidth]{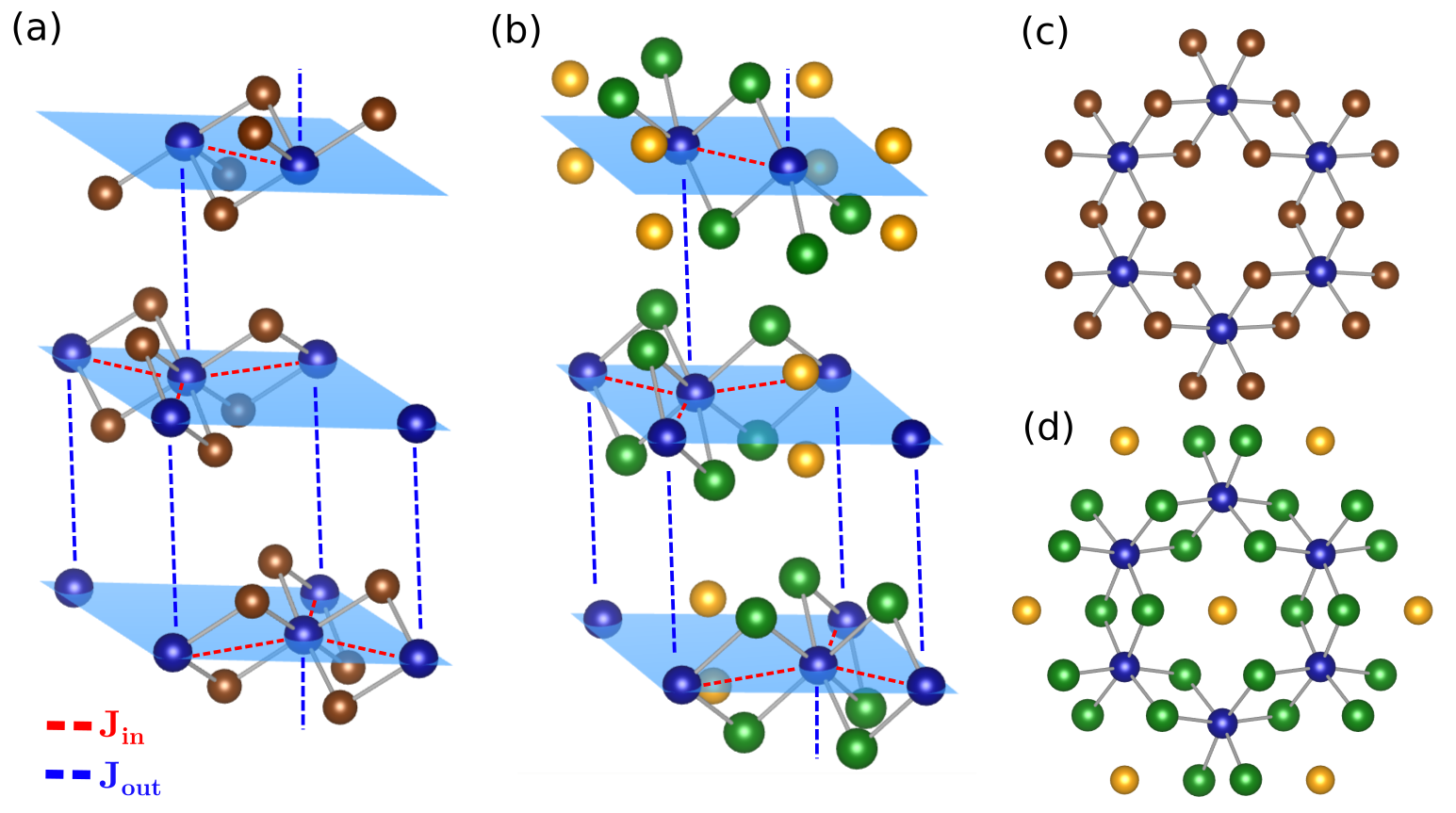}
  \caption{Unit cells of (a) CrBr$_3$ and (b) Cr$_2$Ge$_2$Te$_6$. Cr atom in blue, 
Br atom in brown, Ge atom in yellow and Te atom in green. Different exchange constants are considered: an in-plane 
\textit{J}$_{in}$ that takes into account both the direct Cr-Cr exchange and the superexchange via Br or Te, and the out-of-plane 
\textit{J}$_{out}$ that couples out-of-plane Cr-Cr neighbours. (c) and (d) show the top view of CrBr$_3$ and Cr$_2$Ge$_2$Te$_6$ unit cells, respectively.}
\label{structs}
\end{figure*}

CrBr$_3$ and Cr$_2$Ge$_2$Te$_6$ crystallize in a lamellar structure with hexagonal symmetry (space group no. 148), forming a honeycomb network of edge-sharing octahedra, stacked with van der Waals gaps separating the Cr$^{3+}$-rich planes, as shown in Fig. \ref{structs}. 
Both compounds retain their bulk ferromagnetism down to the single layer limit, T$_C$=34 K for  CrBr$_3$ and T$_C$=61 for Cr$_2$Ge$_2$Te$_6$, with an out-of-plane easy axis.\cite{cr2ge2te6_struct,crbr3_bulk_1,cr2ge2te6_2fm,crbr3_mono}
This magnetic anisotropy is necessary to circumvent the restrictions of the Mermin-Wagner theorem, leading to a long-range FM order at the monolayer limit. 
%Thereby, the evolution of the magnetic anisotropy as a function of the out-of-plane coupling holds the key to understand the stabilization of the 2D magnetic order.

Therefore, the stabilization of 2D magnetic order requires a comprehensive understanding of the magnetic exchange interactions and their evolution as function of order parameters like dimensionality or external pressure. By combining \textit{ab-initio} calculations and high-pressure experiments on single crystals of CrBr$_3$ and Cr$_2$Ge$_2$Te$_6$, we explore their electronic structure, and the pressure dependence of the in- and out-of-plane exchange interactions, \textit{J}$_{in}$ and \textit{J}$_{out}$, (shown in Fig. \ref{structs}). The pressure dependence of the  magnetization measurements shows a different trend of the magnetic transition temperature, T$_C$, for each compound. Our density-functional-theory-based calculations indicate that, while the evolution of T$_C$ at low pressures is driven by progressive decrease of \textit{J}$_{in}$ for both systems, \textit{J}$_{out}$ is the dominant interaction in Cr$_2$Ge$_2$Te$_6$ at high pressure. We argue that the electronic band structure is responsible for the pressure evolution of the magnetic transition temperatures in each compound. Our work highlights the crucial role of electronic band structure to search for new materials retaining a large Curie temperature down to the monolayer limit.

%AÑADIR ALGO DE LA DIRECCION DE MAGNETIZACIÓN VS P CUANDO CALCULADA!!!

The paper is organized as follows. In Section II, we describe the experimental and theoretical approaches used. Section III is devoted to report the electronic structure and magnetic exchange paths of both compounds, Section IV analyses the effect of dimensionality from the evolution of the electronic structure and magnetic properties under pressure. Finally, in Section V we provide a discussion of the results and a summary of the main conclusions of this work.   

\begin{figure*}[!ht]
  \centering
  \includegraphics[width=\textwidth]{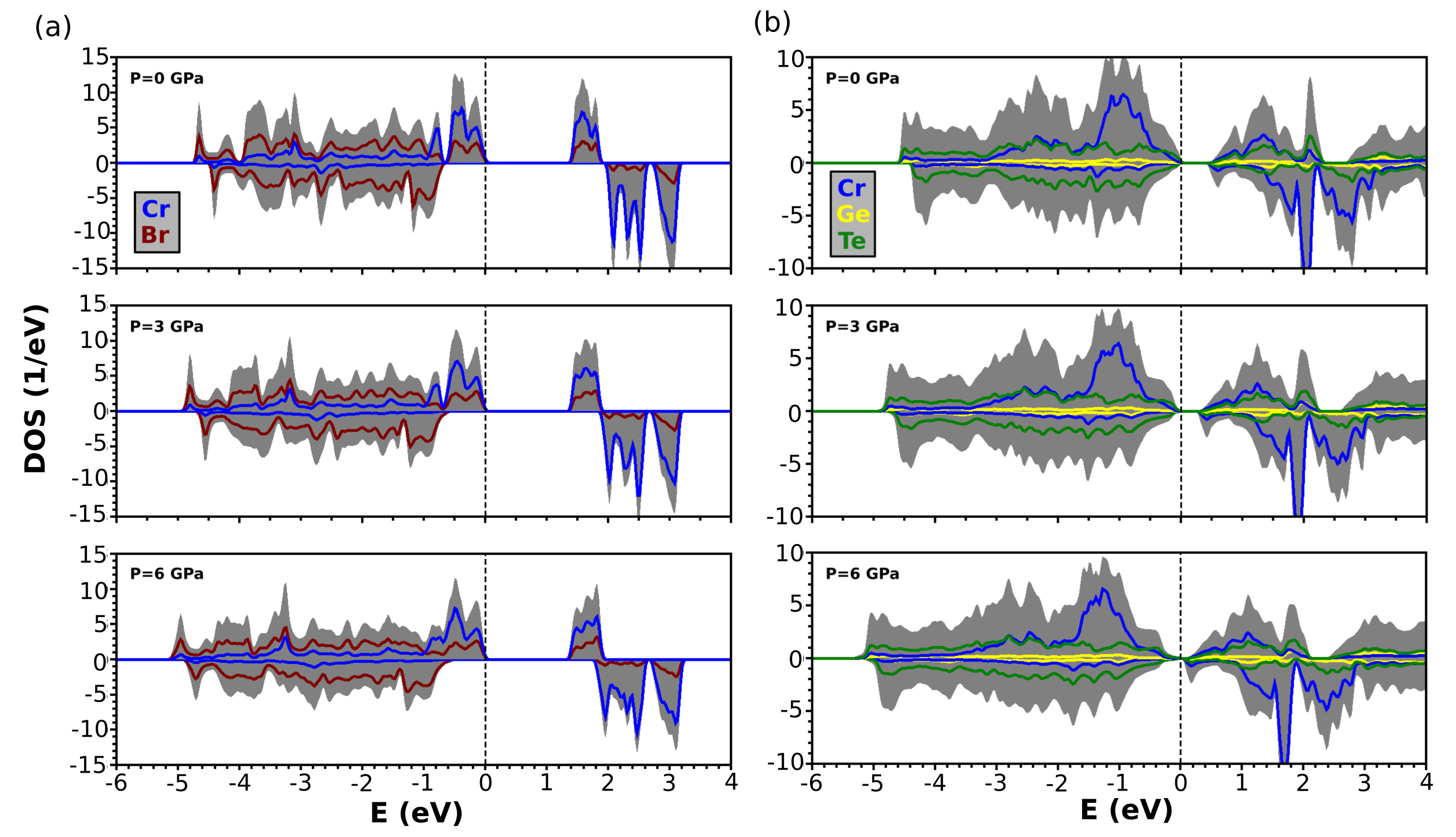}
\caption{ Density of states as a function of pressure. The shaded area represents the total DOS, in positive (negative) the majority (minority) spin channel. Fermi energy was set to zero. (a) Partial density of states (DOS) of the Cr \textit{d}  (blue) and Br \textit{p} (brown) states. The energy gap has \textit{d}-\textit{d} character. (b) Partial density of states (DOS) of the Cr \textit{d}  (blue), Ge \textit{p} (yellow) and Te \textit{p} (green) states. Note the large bandwidth of the Te \textit{p} states leading to a substantial charge transfer; the larger weight at the Fermi level is due to Te \textit{p} bands. The gap closes completely at high pressure in this case.
}
\label{dos}
\end{figure*}

\section{Computational and Experimental Procedures}

Electronic structure \textit{ab-initio} calculations were performed within the density functional theory\cite{dft,dft_2} framework using the all-electron, full potential code {\sc wien2k}\cite{wien}
based on the augmented plane wave plus local orbital (APW+lo) basis set.\cite{sjo} For the structural relaxations, in particular the computation of the lattice parameters at different pressures, special exchange-correlation functionals were needed to deal with the out-of-plane van-der-Waals-type forces.\cite{vanwaals_pot}
The exchange-correlation term chosen to compute the exchange parameters was the generalized gradient
approximation (GGA) in the Perdew-Burke-Ernzerhof scheme.\cite{gga} A fully-converged k-mesh of R$_{mt}$K$_{max}$= 7.0 and muffin-tin radii of  2.42 a.u. for Cr, 2.07 a.u. for Ge, 2.19 a.u. for Br  and 2.38 a.u. for Te, nicely converged with respect to all the input parameters in the simulations. Besides, the Curie temperature (T$_C$) at each pressure was obtained using 864 particles within a nearest-neighbour Heisenberg model, with the exchange parameters mapped by the Monte Carlo Metropolis algorithm.\cite{metropolis} We have applied periodic boundary conditions and $10^6$ steps in total. The T$_C$ was obtained by fitting the magnetization curves as explained in Ref. \cite{Tcfitting}

Single crystals of CrBr$_3$ and Cr$_2$Ge$_2$Te$_6$ were synthesized from the purely elemental starting materials and provided by HQ Graphene. 
The quality of the crystals was confirmed by X-ray diffraction and X-ray photoemission spectroscopy (XPS). X-ray absorption (XAS) and X-ray magnetic circular dichroism (XMCD) measurements up to 6 Tesla were performed at Cr \textit{L$_{2,3}$} edge at the BOREAS BL29 beamline at ALBA synchrotron. 
Magnetic measurements under pressure up to P=10 kbar were performed in a Be-Cu cell, using a silicon oil as pressure medium, inside a MPMS SQUID from Quantum Design. The pressure was monitored \textit{in situ} through the superconducting transition of a Sn pressure gauge. X-ray diffraction under pressure was carried out in a diamond anvil cell using 4:1 methanol:ethanol as pressure media. The pressure dependence of the lattice volume up to 50 kbar was fitted to the Birch-Murnaghan equation.

\section{Electronic structure and magnetism}

In this Section we will describe the electronic structure and magnetic interactions for both compounds. 
Figure \ref{dos} shows the density of states computed with DFT for CrBr$_3$ and Cr$_2$Ge$_2$Te$_6$. 
In both cases, the electronic structure of Cr can be well described as a Cr$^{3+}$:\textit{d}$^3$ cation with \textit{S}=3/2, the majority-spin \textit{t}$_{2g}$ bands being fully occupied. 

On the one hand, for CrBr$_3$, the states just below and above the Fermi level have dominant Cr \textit{d} character, thus suggesting a Mott insulator behaviour with a \textit{d}-\textit{d} energy gap (top panel in Fig. \ref{dos}a). This result is in agreement with previous photoluminescence experiments.\cite{crbr3_mono}

On the other hand, for Cr$_2$Ge$_2$Te$_6$, the states just below the Fermi level are very broad and have dominant Te \textit{p} character. A \textit{p}-\textit{d} energy gap is observed for this material, which suggests a charge transfer insulator behaviour. Ge atoms do not play a substantial role in the electronic structure, with a vanishing contribution close to the Fermi level  (top panel in Fig. \ref{dos}b). 
%This DFT prediction of a charge transfer insulator behaviour for Cr$_2$Ge$_2$Te$_6$ is confirmed by the subsequent experiments.

\begin{figure}[h!]
  \centering
  \includegraphics[width=\columnwidth]{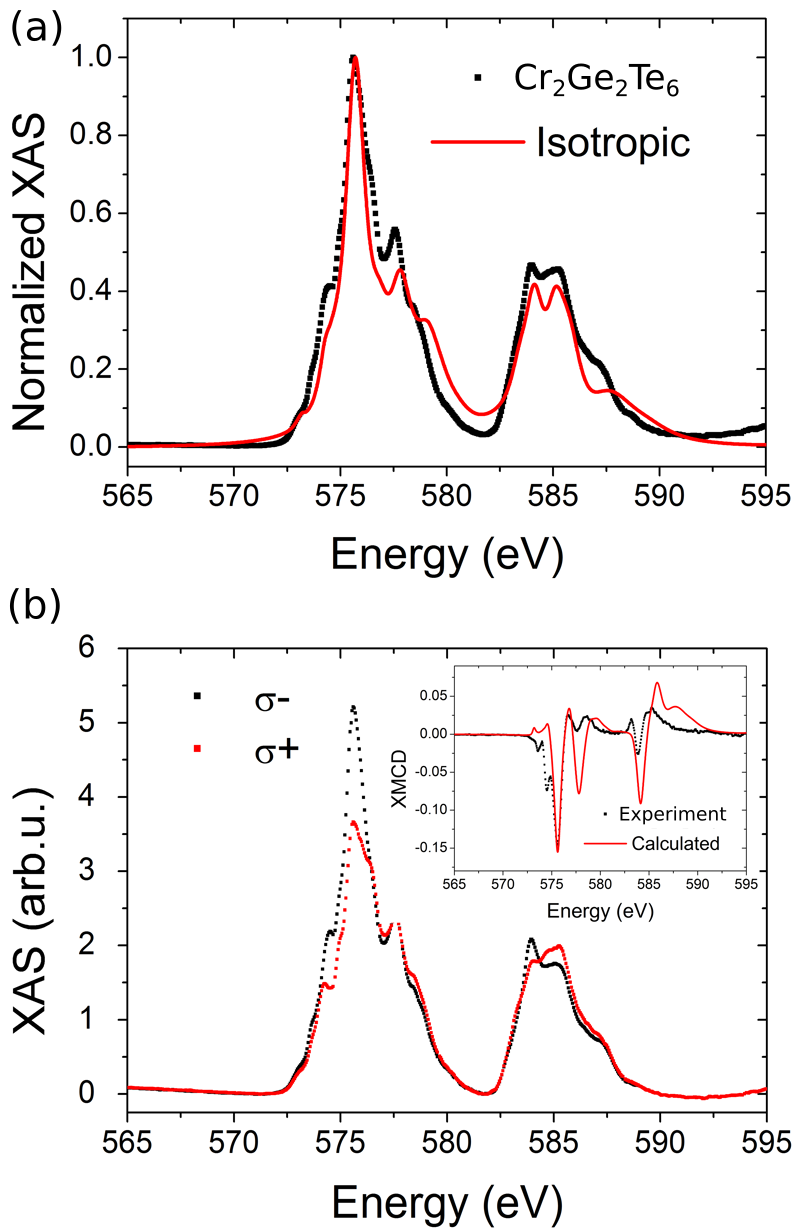}
\caption{(a) X-ray absorption of Cr$_2$Ge$_2$Te$_6$ at the Cr $L$-edge (black dots) and the calculated XAS spectra. (b) Experimental circular left (black) and right (red) polarized light. Inset, experimental and theoretical XMCD spectra for Cr$^{3+}$ with \textit{C}$_{3v}$ symmetry. 
}
\label{spectra}
\end{figure}

Figure \ref{spectra}a shows the experimental isotropic XAS taken at 2 K for Cr$_2$Ge$_2$Te$_6$. The isotropic spectrum is defined as the average of circular $\sigma^-$ and circular $\sigma^{+}$ polarized light. The two main peaks correspond to the $2p_{3/2}$ (575 eV) and $2p_{1/2}$ (585 eV) components of the $2p$ core level (split by spin-orbit coupling). The energy profile is comparable to Cr$_2$O$_3$, demonstrating a Cr$^{3+}$ oxidation state.\cite{Gau2003}

We have performed cluster calculations for the atomic-like $2p^{6}3d^{3} \to 2p^{5}3d^{4}$ transition for Cr$^{3+}$ (C$_{3v}$ symmetry) using the crystal field theory (CFT) implemented in QUANTY.\cite{Hav12,Lu14} The method accounts for the intraatomic $3d$-$3d$ and $2p$-$3d$ Coulomb and exchange interactions, the atomic $2p$ and $3d$ spin-orbit couplings and the local crystal field parameters $D_q$, $D_{\sigma}$ and $D_{\tau}$. We have adopted the values of $U= 0$ eV, $F^0_{dd}= 0.805$ eV, $F^2_{dd}=15.61$ eV, $F^4_{dd}= 9.78$ eV, $F^2_{pd}= 4.89$ eV, $G^1_{pd}= 2.22$ eV and $G^3_{pd}= 1.26$ eV as input of the Slater integrals. The calculated ground state is nearly four-fold degenerate with spin quantum numbers $S_z \sim$ $\pm3/2$ and $\pm$1/2. The main features of the spectra are well modelled with cluster calculations despite the low symmetry of the Cr site and the mixing of \textit{L}$_3$ and \textit{L}$_2$ edges, assuming $D_q\sim 0.3$ eV, $D_{\sigma} = 0$ eV and $D_{\tau}= 0$ eV, therefore, indicating no distortion from the trigonal $C_{3v}$ symmetry.

In Figure \ref{spectra}b, we show the $\sigma^+$ and $\sigma^-$ polarized light and the XMCD spectra (inset) defined as $\sigma^-$-$\sigma^+$ together with the theoretical calculation. Again, the fitting reproduces the main peaks and splittings of the spectra, although the calculation does not reproduce the low energy tail ($\sim$ 572 eV) of the XMCD spectrum (inset). Incident photon angular dependence XMCD (not shown) confirms the out-of-plane magnetic anisotropy. 

\begin{figure}[h!]
  \centering
  \includegraphics[width=\columnwidth]{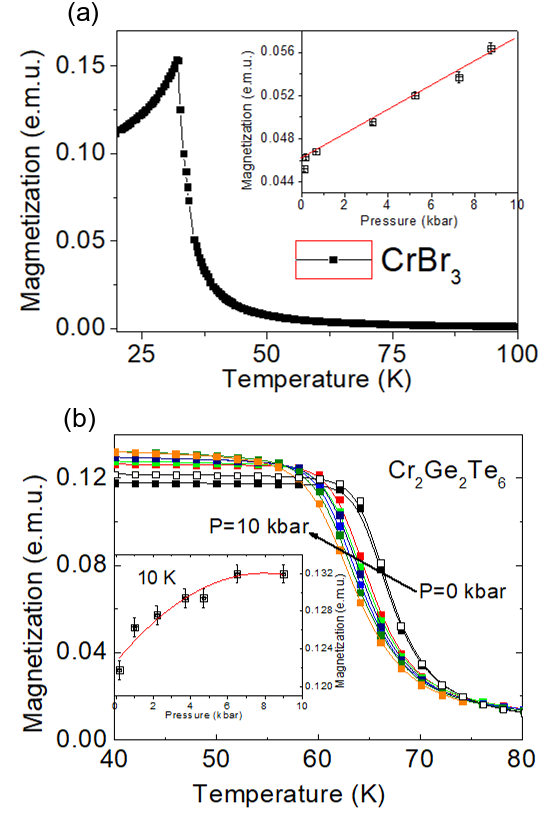}
\caption{(a) Magnetization vs temperature at zero pressure for CrBr$_3$, \textit{H}=0.1 T. Inset, linear pressure dependence of the magnetization at 10 K.  (b) Temperature dependence of the magnetization (\textit{H}=0.1 T) up to 10 kbar for Cr$_2$Ge$_2$Te$_6$. Inset, Pressure dependence of the magnetization. At 10 K and above 6 kbar the magnetization saturates in agreement with the pressure dependence of the \textit{in} and \textit{out} exchange interactions, \textit{J}$_{in,out}$, see text. (red lines are guides to eye).}
\label{fran_0}
\end{figure}

From the theoretical point of view, the different contributions to the magnetic ground state can be modelled separately by considering the nearest-neighbour \textit{J}$_{in}$ and the \textit{J}$_{out}$ exchange couplings. Four more exchange paths including second and third nearest neighbours have been included in previous works.\cite{gong_discovery_2017} Nevertheless, as we will discuss, our simple Heisenberg-type model suffices to explain the magnetic order in these compounds. The different \textit{J}'s are shown in Fig. \ref{structs}, with the \textit{J}$_{in}$ and \textit{J}$_{out}$ depicted as broken red and blue lines, and the Cr$^{3+}$ cations surrounded by edge-sharing trigonally distorted octahedra. \textit{J}$_{in}$ is the sum of two exchange contributions, the direct Cr-Cr exchange across the octahedral edge, which is antiferromagnetic (AF) between two half-filled \textit{t}$_{2g}$ bands, and the indirect Cr-(Br or Te)-Cr superexchange at approximately 90-degrees. According to Goodenough-Kanamori-Anderson rules, the indirect exchange, cation-anion-cation, can be either FM or AF, depending on the strength of the delocalization and correlation superexchange effects.\cite{goody} Our results show that \textit{J}$_{in}=90$ K and \textit{J}$_{out}=12$ K for Cr$_2$Ge$_2$Te$_6$ and \textit{J}$_{in}=89$ K and \textit{J}$_{out}=6$ K for CrBr$_3$. These values were obtained by fitting total energy differences between different magnetic configurations to a Heisenberg-type Hamiltonian. All \textit{J}'s are positive, which explains the FM order observed in the experiments for both compounds. Figure \ref{fran_0} shows the magnetization \textit{vs} temperature at at P$=0$ GPa for CrBr$_3$ and up to 10 kbar for Cr$_2$Ge$_2$Te$_6$. The magnetic transitions appear as an upturn in the magnetization upon cooling at 37  and 63 K, respectively, following a Curie-Weiss behavior at T$>$T$_C$.

\section{Evolution with pressure}
We have carried out \textit{ab-initio} calculations, complemented with high-pressure magnetic and X-ray diffraction measurements to study the evolution of the electronic structure and magnetic exchange interactions. Naively, one can interpret the effect of isotropic pressure on the magnetic exchange interactions as a smooth evolution from a 2D- toward a more 3D-like magnetic state, since the \textit{c}-axis compressibility is expected to be much larger than that of the \textit{a-b} plane.

\begin{figure}[h!]
  \centering
  \includegraphics[width=\columnwidth]{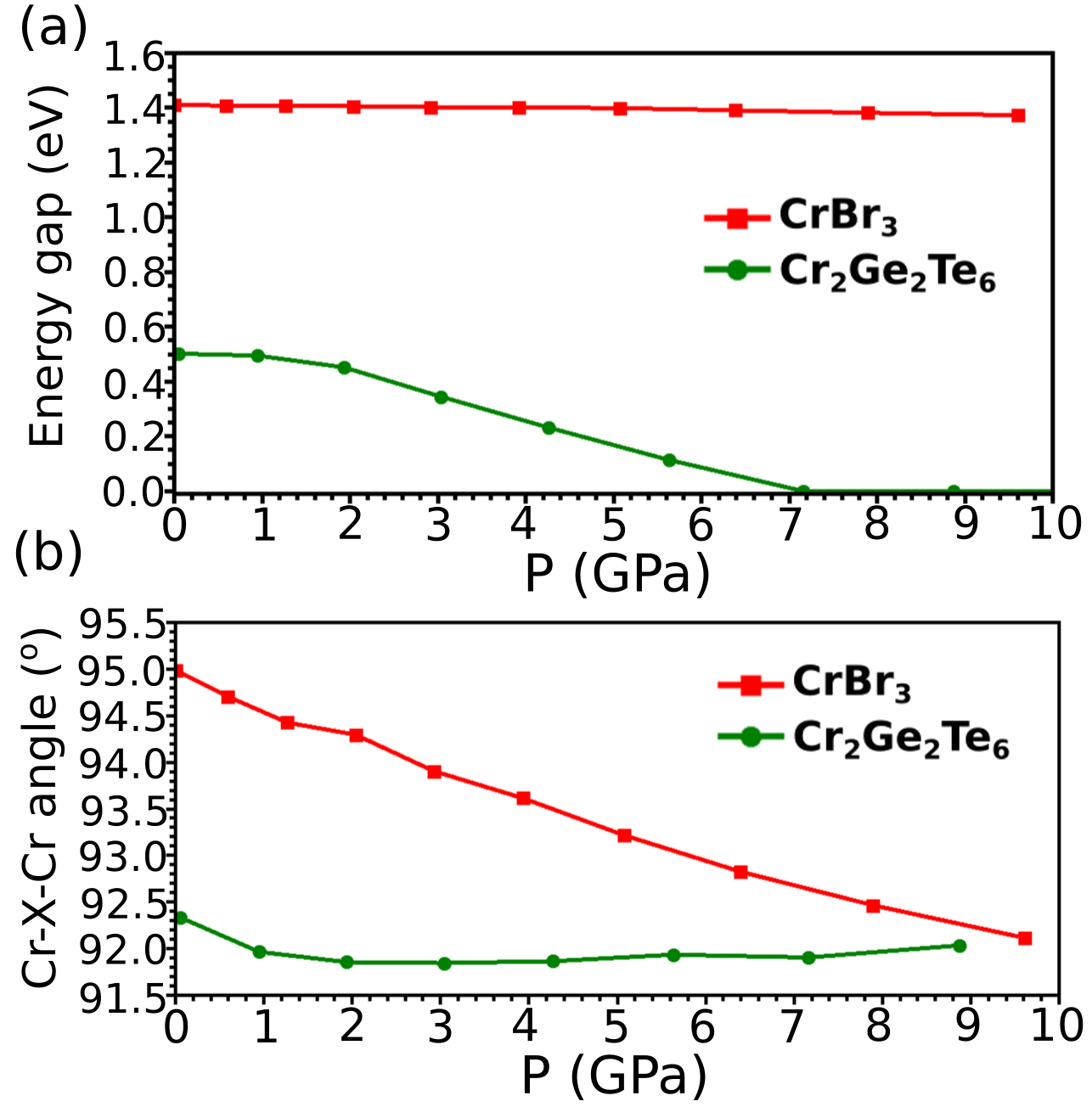}
\caption{(a) Calculated energy gap as a function of pressure for both compounds, obtained using the GGA exchange-correlation functional. (b) Evolution of the Cr-X-Cr angle as a function of pressure. X represents Br or Te atoms. 
}
\label{elecpres}
\end{figure}

Figure \ref{dos} shows the DOS and the partial DOS (pDOS) as a function of pressure for both compounds. In Fig. \ref{dos}a we see that CrBr$_3$ does not undergo a substantial modification of its electronic structure with pressure. 
The \textit{d}-\textit{d} energy gap observed at P=0 GPa is preserved when pressure is raised up to 6 GPa (see Fig. \ref{elecpres}a). However, Figure \ref{dos}b shows how the \textit{p}-\textit{d} energy gap of Cr$_2$Ge$_2$Te$_6$ closes as pressure is increased (see Fig. \ref{elecpres}a). Therefore, our calculations predict that Cr$_2$Ge$_2$Te$_6$ undergoes an insulator to metal transition at high pressure. The values of the band gaps presented here were obtained using GGA, which has a tendency to underestimate them. Yet, the pressure evolution should be correct.

Figure \ref{elecpres}b shows the pressure dependence of the cation-anion-cation angle. While for the Mott insulator CrBr$_3$ the angle shows a monotonic decrease up to 10 GPa, in Cr$_2$Ge$_2$Te$_6$ the angle remains constant in the whole range of pressures up to 10 GPa. Hence, external pressure modifies the internal structure of CrBr$_3$, while in Cr$_2$Ge$_2$Te$_6$, there is a pressure induced insulator to metal transition as a consequence of the closing of the energy gap due to the change in the Te-Cr distance and not the Cr-Te-Cr angle, which remains roughly unchanged.

\begin{figure}[h!]
  \centering
  \includegraphics[width=\columnwidth]{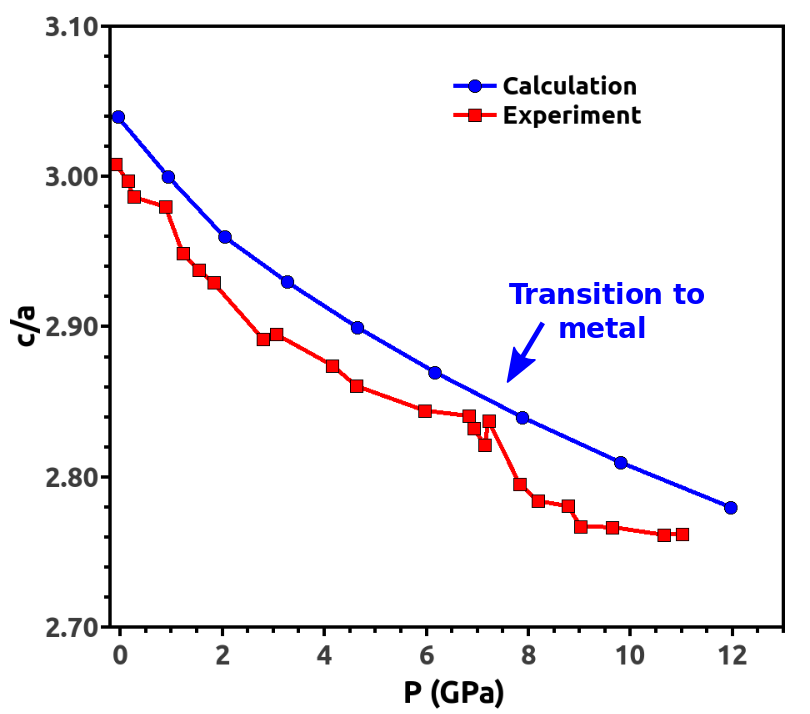}
\caption{Comparison of the high pressure XRD data with the results of our ab initio calculation for the c/a ratio. DFT predicts a transition to a metallic state around the same pressure that XRD data shows a peak.
}
\label{xrd}
\end{figure}

In order to shed light on the nature of the pressure-induced metal-insulator transition predicted by our DFT calculations for Cr$_2$Ge$_2$Te$_6$, we have performed high pressure X-ray diffraction (XRD) measurements. In Fig. \ref{xrd}, we plot the pressure dependence of the ratio of the lattice parameters, $c/a$, showing a sudden decrease above 7 GPa, suggesting a structural transition. This pressure coincides with the collapse of the electronic gap, as predicted by our \textit{ab initio} calculations. 
The insulator to metal transition observed in our DFT calculations was obtained assuming that the symmetry of the lattice remains unchanged upon pressure. Pressure-induced structural phase transitions and amorphizations have been observed before by means of high pressure X-ray diffraction between 18 and 30 GPa for Cr$_2$Ge$_2$Te$_6$. However, no structural anomalies were reported at pressures relevant in this work.\cite{yu2019pressure}

\begin{figure}[h!]
  \centering
  \includegraphics[width=\columnwidth]{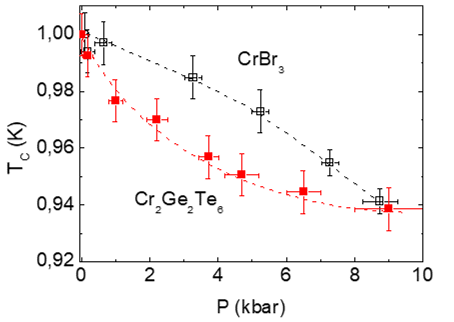}
\caption{(a) Summary of the pressure dependence of the normalized Curie temperature, T$_C$ for Cr$_2$Ge$_2$Te$_6$ and CrBr$_3$.
}
\label{fran_p}
\end{figure}

\begin{figure}[h!]
  \centering
  \includegraphics[width=\columnwidth]{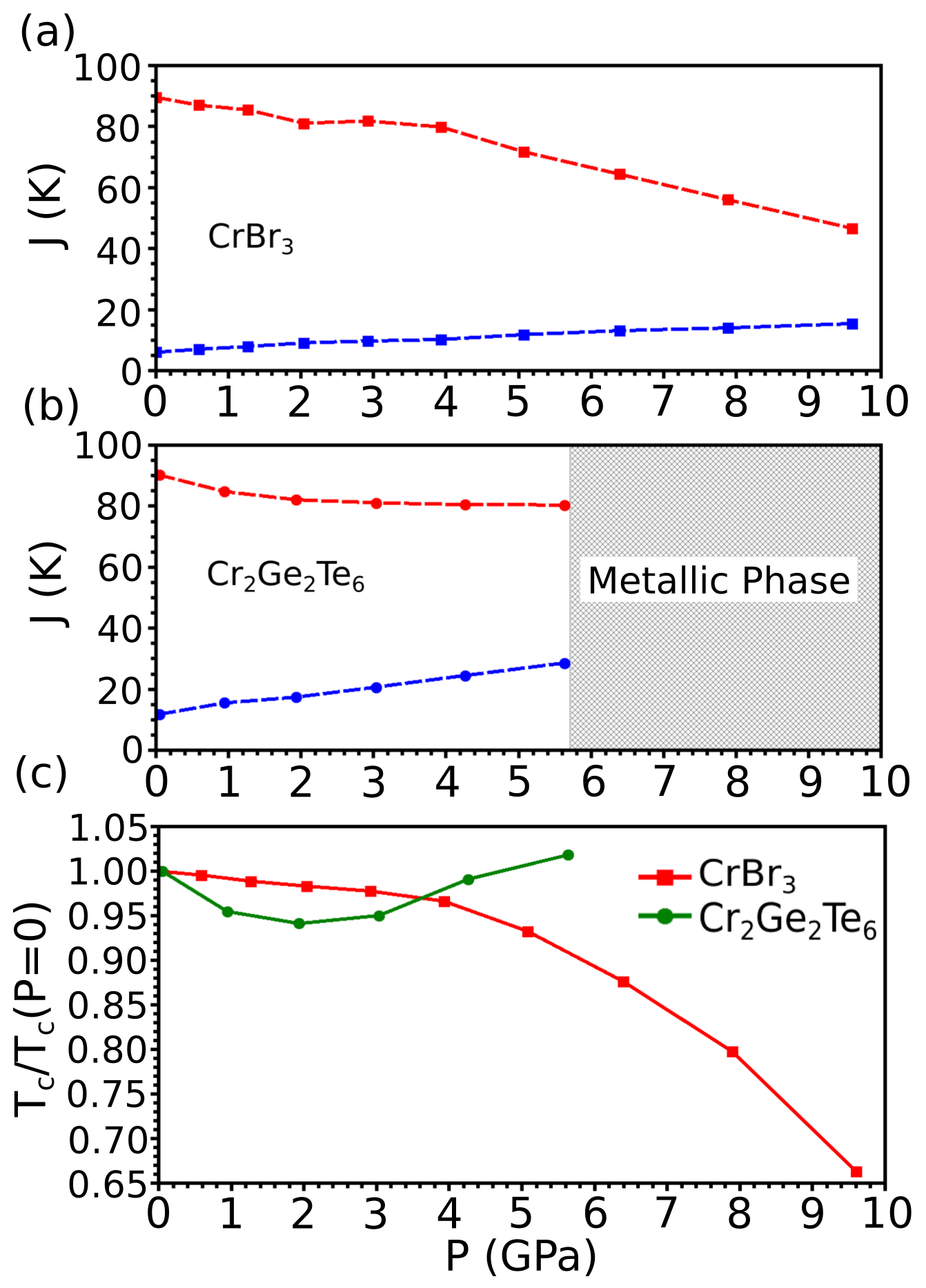}
\caption{(a) and (b) values of the exchange constants extracted using a Heisenberg-type Hamiltonian for CrBr$_3$ and for Cr$_2$Ge$_2$Te$_6$, respectively. J$_{out}$ in blue, J$_{in}$ in red. The out-of-plane coupling becomes larger as pressure is applied. (c) Evolution of the normalized transition temperature with pressure obtained from the Monte-Carlo simulation.}
\label{J_Tc}
\end{figure}

Figures \ref{fran_0} and \ref{fran_p} display the evolution of the magnetic transition temperature of CrBr$_3$ and Cr$_2$Ge$_2$Te$_6$ up to 1GPa. As seen in the raw data and summarized in  Fig. \ref{fran_p}, the T$_C$ of both compounds decreases as pressure is applied, in agreement with previous reports\cite{YOSHIDA1997525,crgete_pressure} and at odds with the expectations for the pressure dependence of the exchange interaction in a localized magnetic system. Harrison\cite{Harrison} stated that the transition temperature for a localized system increases as a function of the distance between magnetic sites as \textit{r}$^{-(l+l'+1)}$, where \textit{l} and \textit{l'} are the angular momentum quantum numbers. This results in a variation of the magnetic exchange interaction with volume, \textit{J(V)} as 3.3 and 4.6 for direct and indirect superexchange, respectively, the so-called Bloch's rule \cite{Bloch}. Violations of  Bloch's rule have been observed in Mott insulators close to a metal-insulator \cite{Blanco2012} and spin-Peierls to Peierls \cite{Blanco2013} transitions. Bloch's rule describes the evolution with pressure of only one exchange constant. As we have seen above, in this case there is a competition between exchange constants that evolve differently with pressure, and even the J$_in$ is in itself the addition of two competing exchange mechanisms.
Our data also shows a different trend for T$_C$(P) than the simple Bloch's rule prediction: while the critical temperature of CrBr$_3$ displays a slightly convex decay, the decrease observed in Cr$_2$Ge$_2$Te$_6$ is concave and shows a clear tendency to saturation at the highest pressures analyzed in this work. This behavior seems to go in hand with the pressure dependence of the magnetization at 10 K, which increases linearly with pressure for CrBr$_3$ (inset of Figure \ref{fran_0}a), but remains nearly constant above 6 kbar for Cr$_2$Ge$_2$Te$_6$, inset of Figure \ref{fran_0}b. 

%\color{red}{Indeed, an enhancement of T$_C$ and a decrease of magnetization under pressure has been reported in CrI$_3$ and discussed within the pressure dependence of the coupling between the layers and Cr-I-Cr bond angle. Phys. Rev. B 99, 180407(R) – Published 13 May 2019. In addition, the magnetic transition seems to be unaffected by pressure up to 0.6 GPa in VI$_3$, followed by a dramatic enhancement of T$_C$ up to 1 GPa. PHYSICAL REVIEW B 99, 041402(R) (2019), presumably due to the correlated character of the Mott insulator.

As previously described, the electronic structure shows largely ionic Cr$^{3+}$:S$=3/2$ cations with a half-filled spin-polarized \textit{t}$_{2g}$ and \textit{e}$_g$ manifold. This leads to a significant gap opening at the Fermi level for Cr$_2$Ge$_2$Te$_6$ and CrBr$_3$, even at the LDA/GGA level, without the need to introduce strong correlation effects in the calculations. In a first approximation, this indicates that the magnetic semiconductor limit could be applicable to understand the pressure dependence of these systems.  Note that for the case of Cr$_2$Ge$_2$Te$_6$ we have only performed this analysis at pressures where the system remains a semiconductor. 
However, even if the system can be described in the localized limit, various different exchange interactions compete, responding differently to pressure, and this leads to a complex pressure dependence of T$_C$. We have calculated the pressure dependence of both \textit{J}$_{in}$ and \textit{J}$_{out}$. 
The system is fully relaxed at different volumes (including the lattice parameters \textit{c}/\textit{a} ratio and the internal positions), and used the Birch-Murnaghan equation for obtaining the pressure value at each volume.\cite{birch_murnaghan}
We set up a  magnetic supercell combining different types of AF and FM couplings between the neighbouring Cr cations. The value of the magnetic exchange constants is obtained after subtraction of the total energy difference among various magnetic configurations, assuming these follow a simplistic Heisenberg-type spin Hamiltonian with neighbouring 3/2 spins for Cr as explained before. The results are summarized in Fig. \ref{J_Tc}. The ground state presents a FM coupling at low pressures, in agreement with experiments, but the different exchange constants behave differently with pressure, although all of them remain positive at all pressures, for both compounds.

With the calculated values of the magnetic exchange constants, we have set up a Heisenberg type Hamiltonian of interacting spins, and solved it using a classical Monte Carlo simulation. This allowed us to obtain a theoretical T$_C$ and its evolution with pressure. Despite  the overestimation of the absolute value of magnetic exchange constants, the actual pressure dependence of the calculated T$_C$ is in good agreement with the observed experimental trend, as it can be seen from a comparison between Figs. \ref{fran_p}a and \ref{J_Tc}c. In this figure, the theoretical values are normalized to the experimental T$_C$ at ambient pressure. In the Heisenberg type Hamiltonian we have included an anisotropic term as in ref. \cite{gong_discovery_2017}, since it was reported that for Cr$_2$Ge$_2$Te$_6$ this term changes its sign with pressure.\cite{spin_anys_press} However, no influence on the value of T$_C$ or its pressure dependence was observed. 

For CrBr$_3$, \textit{J}$_{out}$ increases with pressure while \textit{J}$_{in}$ decreases (Fig. \ref{J_Tc}a). Since \textit{J}$_{in}$ $>$ \textit{J}$_{out}$ and the number of Cr in-plane neighbours is greater than the out-of-plane, the trend of T$_C$(P) is governed by \textit{J}$_{in}$. Therefore, decrease of \textit{J}$_{in}$ produces the decrease of T$_C$ as pressure increases, in agreement with the experimental values in Figure \ref{fran_p}. On the other hand, for Cr$_2$Ge$_2$Te$_6$, \textit{J}$_{out}$ increases with pressure, while, initially, \textit{J}$_{in}$ decreases and saturates at higher pressure (see Fig. \ref{J_Tc}b). Therefore, \textit{J}$_{in}$ dominates the change of T$_C$ at low pressures, but \textit{J}$_{out}$ becomes the dominant term at higher pressures. This is in agreement with the experimental evolution presented for the T$_C$ with pressure (fig. \ref{fran_p}). 
Further experimental measurements at higher pressures should confirm the minimum in T$_C$ predicted in our calculations.

\section{Conclusions}\label{summary}

To summarize, pressure effects on the electronic,  magnetic and structural properties of CrBr$_3$ and Cr$_2$Ge$_2$Te$_6$ have been explored following a combination of experimental and computational approaches. 
Our results show that CrBr$_3$ is a Mott-Hubbard like insulator with a \textit{d}-\textit{d} energy gap. It follows that the system remains insulating under pressure. On the other hand, the energy gap of Cr$_2$Ge$_2$Te$_6$ is found to have \textit{p}-\textit{d} character, which leads to a charge transfer insulating behaviour. Our calculations predict an insulator to metal transition to occur at 6 GPa assuming no structural transition takes place. Despite the high pressure X-ray data shows a structural transition at similar pressures, whether this structural transition is accompanied with a softening of the lattice on approaching the putative insulator to metal transition or a mere structural change to a lower symmetry semiconducting state, cannot be concluded from our data. Additionally resistivity measurements under pressure are needed to clarify this issue. 

Besides, we have found that the different electronic structure of CrBr$_3$ and Cr$_2$Ge$_2$Te$_6$ and its evolution with pressure are at the root of the pressure dependence of the Curie temperature and the magnetic exchange interactions in Cr$_2$Ge$_2$Te$_6$ and CrBr$_3$.
The experimental observation yields a reduction of the Curie temperature as pressure is applied, in contrast to isostrucutral CrI$_3$ and VI$_3$, but with a different trend for each compound. \emph{Ab initio} calculations reveal that the microscopic origin for this trend relies on the different pressure dependence of the in- and out-of-plane magnetic exchange couplings, \textit{J$_{in,out}$}$>$0.

%\color{red}{However, the in-plane magnetic exchange coupling constants are found to be pressure-dependent on the electronic structure, while the out-of-plane constant increases its strength (ferromagnetic character) with pressure for both compounds. Non se ben entende esto. Se pode reformular de outra maniera?}.  

%The fact that Cr$_2$Ge$_2$Te$_6$ is closer to a metal-insulator transition than the more ionic CrBr$_3$ is reflected in the magnetic, structural and transport properties of the system, making them substantially different. 
While the transition temperature for CrBr$_3$ is expected to monotonically decrease with pressure, the proximity of Cr$_2$Ge$_2$Te$_6$ to the metallic limit yields a non-monotonic behavior. This effect was explained as a different behaviour of the in-plane magnetic exchange coupling constants as a function of pressure for each compound.

This work sheds light on the competition of the different types of exchanges that may occur in several other quasi-two-dimensional magnetic materials. Similar competition between direct and indirect in-plane exchange paths and off-plane couplings may occur, such as the di- and trihalides. In fact, it might be at the origin of the presence or absence of magnetic ordering in the monolayer of transition metal phosphorus trisulfides (TMPS$_3$) FePS$_3$ and NiPS$_3$\cite{lee2016ising,kim2019suppression}. Long-range order depends on the type of spin-spin interations, which themselves compete with intrinsic fluctuations and determine the magnetic anisotropy and the stability of long-range magnetic order at the monolayer limit. We have seen that in Cr-based van der Waals structures, these spin-spin interactions (and their competitions) are strongly dependent on the evolution of the electronic structure with pressure and, therefore reciprocally, on the  dimensionality.
%\color{black}{

\section*{Conflicts of interest}

There are no conflicts to declare.

\section*{Acknowledgments}

This work was supported by the Spanish Government through Projects MAT2016-80762-R,  PGC2018-101334-B-C21 and PGC2018-101334-A-C22. A.O.F. also through the FPU16/02572 grant. J.S.Z was supported by National Science Foundation grant DMR-1720595.

\end{document}